\documentclass[11pt,amsmath,prl,preprint,showpacs,superscriptaddress]{revtex4}
\usepackage{graphicx}

\begin{document}

\title{Current induced light emission and light induced current\\ 
       in molecular tunneling junctions}

\date{\today}

\author{Michael Galperin}
\altaffiliation[Present address:]{Department of Chemistry, Nortwestern University, Evanston IL 60208, U.S.A.}
\author{Abraham Nitzan}
\affiliation{School of Chemistry, Tel Aviv University, Tel Aviv 69978, Israel}
      
\begin{abstract}
The interaction of metal-molecule-metal junctions with light is considered 
within a simple generic model. We show, for the first time, that light induced 
current in unbiased junctions can take place when the bridging molecule 
is characterized by a strong charge-transfer transition. The same model shows 
current induced light emission under potential bias that exceeds the molecular 
excitation energy. Results based on realistic estimates of molecular-lead 
coupling and molecule-radiation field interaction suggest that both effects 
should be observable.
\end{abstract}


\maketitle

Molecular conduction nano-junctions have been under intense study for some 
time. A class of molecules not yet investigated in this context are those 
characterized by strong charge-transfer transitions into their first excited 
state. The dipole moment of such molecules changes considerably in such
transitions\cite{1ss}, expressing a strong shift of the electronic charge 
distribution. In the independent electron picture, the charge transfer nature 
of the first excited state implies that either the highest occupied, or the 
lowest unoccupied, molecular orbitals (LUMO and HOMO) is dominated by atomic 
orbitals of larger amplitude (and better overlap with metal orbitals) on one 
side of the molecule than on the other and therefore stronger coupling to one 
of the leads. We show that when such molecular wire connects between two metal 
leads with the charge transfer direction approximately parallel to its axis, 
optical pumping may create an internal driving force for charge flow between 
the leads.  Such optical-resonance activation 
of current flow is different from previously demonstrated adiabatic 
pumping\cite{2ss} and from previously discussed strong field optical control 
mechanisms\cite{3ss,4ss}.

The opposite phenomenon: light emission in current carrying tunnel junctions 
has already been demonstrated. Emission from bare junction\cite{5ss} was 
attributed to surface plasmons\cite{6ss}. Emission from molecular 
nanojunctions\cite{7ss,8ss} was not so far discussed theoretically.

Our model provides a simple framework for treating both light induced current 
and current induced light in molecular junctions. Following its introduction 
below we present results of self consistent numerical solutions as well as 
approximate expressions for the corresponding electron and photon fluxes. 
We examine, using reasonable experimentally-based parameters, the magnitudes 
of these effects and conclude that both phenomena can be observable 
under fairly general conditions.

\begin{figure}[htbp]
\includegraphics[width=\columnwidth]{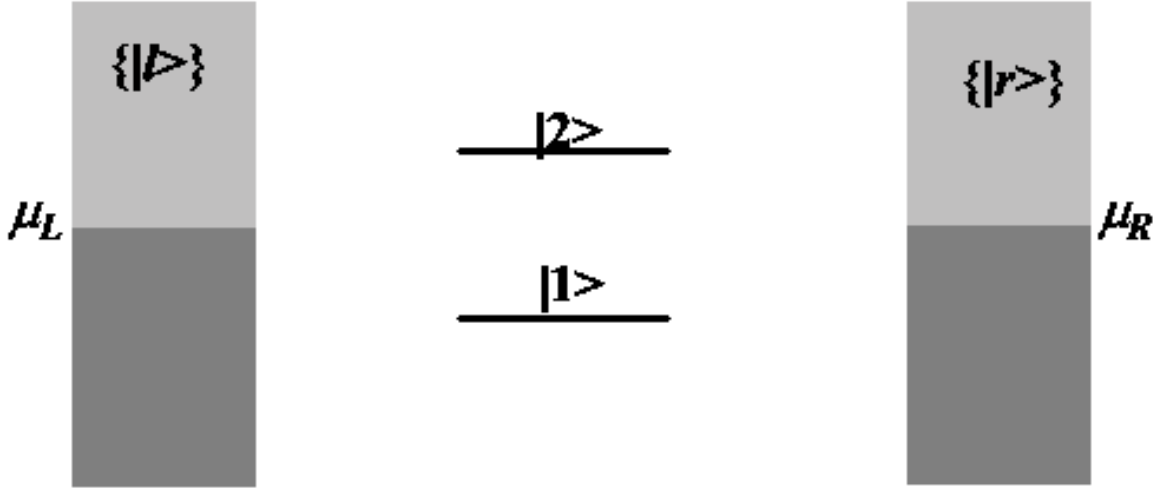}
\caption{\label{Fig1}The molecular junction model. The right ($R=\{|r>\}$) 
and left ($L=\{|l>\}$) manifolds represent the leads' electronic states;
$\mu_L$ and $\mu_R$ are the corresponding electrochemical potentials.
The molecule is represented by its HOMO and LUMO levels.}
\end{figure}
      
In our model (Fig.~\ref{Fig1}) the molecule is represented by its HOMO, $|1>$, 
and LUMO, $|2>$, with energies $\varepsilon_1$ and $\varepsilon_2$ and gap 
$\varepsilon_{21}=\varepsilon_2-\varepsilon_1$, positioned between two leads 
represented by free electron reservoirs $L$ and $R$ characterized by the 
electronic chemical potentials $\mu_L$ and $\mu_R$. $\mu_L-\mu_R=e\Phi$ is 
the voltage bias. In the independent electron picture molecular excitation
corresponds to transfer of an electron between levels $|1>$ and $|2>$. 
The corresponding Hamiltonian is
\begin{align}
 \label{Eq1ss}
 &\hat H = \hat H_0 + \hat V_M + \hat V_N + \hat V_P 
         = \hat H_J + \hat V_N + \hat V_P
 \\
 \label{Eq2ss}
 &\hat H_0 = \sum_{m=1,2}\varepsilon_m\hat c_m^\dagger\hat c_m
           +  \sum_{k\in\{L,R\}}\varepsilon_k\hat c_k^\dagger\hat c_k
           + \hbar\sum_\alpha \omega_\alpha\hat a_\alpha^\dagger\hat a_\alpha \\
 \label{Eq3ss}
 &\hat V_M = \sum_{K=L,R}\sum_{m=1,2;k\in K}\left(
              V_{km}^{(MK)}\hat c_k^\dagger\hat c_m+\text{H.c.}\right) \\
 \label{Eq4ss}
 &\hat V_N = \sum_{K=L,R}\sum_{k\neq k'\in K}\left(
              V_{kk'^{(NK)}\hat c_k^\dagger\hat c_{k'}\hat c_2^\dagger\hat c_1}
              +\text{H.c.}\right) \\
 \label{Eq5ss}
 &\hat V_P = 
 \begin{cases}
              \left(V_0^{(P)}\hat a_0\hat c_2^\dagger\hat c_1
              +\text{H.c.}\right) & \text{case a}\\ 
              \sum_\alpha\left(V_\alpha^{(P)}\hat a_\alpha
              \hat c_2^\dagger\hat c_1+\text{H.c.}\right) & \text{case b}
 \end{cases}
\end{align}
where $\text{H.c.}$ denotes Hermitian conjugate. $\hat H_0$ in (\ref{Eq2ss}) 
contains additively terms that correspond to the isolated molecule, the free 
leads and the radiation field. Eqs.~(\ref{Eq3ss})-(\ref{Eq5ss}) describe 
the coupling between these subsystems. The operators $\hat c$ ($\hat c^\dagger$)
and $\hat a$ ($\hat a^\dagger$) are annihilation (creation) operators for 
electrons and photons. $\hat V_M$ is the standard electron 
transfer coupling that gives rise to net current in the biased junction and 
$\hat V_N$ describes energy transfer between the molecule and electron-hole 
excitations in the leads. The latter interaction strongly affects the
lifetime of excited molecules near metal surfaces and should be inlcuded
in any analysis of phenomenainvolving resonance radiation and molecular 
conduction junctions. The molecule-radiation field coupling $\hat V_P$, 
Eq.~(\ref{Eq5ss}), will be taken in two forms: The form (\ref{Eq5ss}a) 
describes driving of the junction by the electromagnetic field mode $\alpha=0$, 
Eq.~(\ref{Eq5ss}b) is used when addressing the issue of spontaneous light 
emission from current carrying junctions. We limit ourselves to near resonance 
processes pertaining to linear spectroscopy. This justifies the use of the 
rotating wave approximation (RWA) in Eq.~(\ref{Eq5ss}). Note that the radiative 
coupling coefficients  reflect properties of the local electromagnetic field at 
the molecular bridge which depend on the metallic boundary conditions. 

In the Keldysh non-equilibrium Green function approach the steady state
currents through the bridge are obtained from
\begin{equation}
 \label{Eq6ss}
 I_B = \int_{-\infty}^{+\infty}\frac{dE}{2\pi\hbar}\text{Tr}
 \left[\mathbf{\Sigma_B^{<}}(E)\mathbf{G^{>}}(E)
      -\mathbf{\Sigma_B^{>}}(E)\mathbf{G^{<}}(E)\right] 
\end{equation}
were the Green functions (GFs) $G$ and the self energies (SEs) $\Sigma$ are 
defined in the bridge subspace. The subscript $B$ corresponds to a particular 
relaxation processes represented by the SE $\Sigma_B^{<,>}$. At steady state 
the absorption and emission photon fluxes, $I_{abs}$ and $I_{em}$, the 
non radiative relaxation flux, $I_N$, and the source-drain current, $I_{sd}$, 
come into balance. The SEs needed for their evaluation are calculated as sums 
of independent contributions associated with the different relaxation 
processes (the non-crossing approximation)
\begin{equation}
 \label{Eq7ss}
 \mathbf{\Sigma} = \mathbf{\Sigma_{ML}} + \mathbf{\Sigma_{MR}} 
                 + \mathbf{\Sigma_{P}} 
                 + \mathbf{\Sigma_{NL}} + \mathbf{\Sigma_{NR}}
\end{equation}
where $\Sigma_X$ ($X=ML,MR,P,NL,NR$) is the SE associated with the coupling 
$V^{(X)}$ in Eqs.~(\ref{Eq3ss})-(\ref{Eq5ss}). On the Keldysh contour these 
SEs are $2\times 2$ matrices in the bridge space
\begin{align}
 \label{Eq8ss}
 &\Sigma_{MK,mm'}(\tau_1,\tau_2) = \sum_{k\in K} V_{mk}^{(MK)}
 g_k(\tau_1,\tau_2) V_{km'}^{(MK)} \\
 \label{Eq9ss}
 &\Sigma_{NK,mm'}(\tau_1,\tau_2) = \delta_{mm'}\sum_{k\neq k'\in K}
 \left|V_{kk'}^{(NK)}\right|^2 
 \nonumber\\* &\qquad\times
 g_k(\tau_2,\tau_1)g_{k'}(\tau_1,\tau_2) G_{\bar m\bar m}(\tau_1,\tau_2) \\
 \label{Eq10ss}
 &\mathbf{\Sigma_P}(\tau_1,\tau_2) = i\sum_\alpha\left|V_\alpha^{(P)}\right|^2
 \nonumber \\* &\times
 \left[\begin{array}{cc}
 F_\alpha(\tau_2,\tau_1)G_{22}(\tau_1,\tau_2) & 0 \\
 0 & F_\alpha(\tau_1,\tau_2)G_{11}(\tau_1,\tau_2)
 \end{array}\right]
\end{align}
here and below $K=L,R$ denotes the left and right lieads, $m,m'=1,2$ and
$\bar m=2\delta_{m,1}+\delta_{m,2}$. $g_k$ and $F_\alpha$ are free electron 
and photon GFs in state k and mode $\alpha$, respectively. The 
retarded/advanced, lesser and greater projections of these SEs on the real
time axis are obtained using the Langreth formulas and can be expressed at
steady state situations in energy space. In the wide-band approximation and
assuming that the HOMO and LUMO are not mixed by their interactions with the
leads, the SEs associated with electron exchange between molecule and leads 
have the familiar forms
\begin{subequations}
\label{Eq11ss}
\begin{align}
 \label{Eq11ssa}
 &\Sigma^r_{MK,mm'} = \left[\Sigma^a_{MK,m'm}\right]^{*}
 =-i\delta_{mm'}\Gamma_{MK,m}/2 \\
 \label{Eq11ssb}
 &\Sigma^{<}_{MK,mm'} = i\delta_{mm'}f_K(E)\Gamma_{MK,m} \\
 \label{Eq11ssc}
 &\Sigma^{>}_{MK,mm'} = -i\delta_{mm'}[1-f_K(E)]\Gamma_{MK,m} \\
 \label{Eq11ssd}
 &\Gamma_{MK,m} = 2\pi\sum_{k\in K}\left|V^{(MK)}_{km}\right|^2
                   \delta(E-\varepsilon_k) \\
 \label{Eq11sse}
 &f_K(E) = \left[\exp\left\{(E-\mu_K)/k_BT\right\}+1\right]^{-1}
\end{align}
\end{subequations}
The SEs associated with the non-radiative and radiative energy relaxation 
processes are similarly obtained in the forms
\begin{subequations}
\label{Eq12ss}
\begin{align}
 \label{Eq12ssa}
 &\Sigma_{NK,mm'}^{<}(E) = \delta_{mm'} \int\frac{d\omega}{2\pi}
  B_{NK}(\omega,\mu_K)G_{\bar m\bar m}^{<}(E+\omega) \\
 \label{Eq12ssb}
 &\Sigma_{NK,mm'}^{>}(E) = \delta_{mm'} \int\frac{d\omega}{2\pi}
  B_{NK}(\omega,\mu_K)G_{\bar m\bar m}^{>}(E-\omega) \\
 \label{Eq12ssc}
 &B_{NK}(\omega,\mu_K) = \int\frac{dE}{2\pi}C_{NK}(E,\omega)
                          f_K(E)[1-f_K(E+\omega)] \\
 \label{Eq12ssd}
 &C_{NK}(E,\omega) = \left(2\pi\right)^2\sum_{k\neq k'\in K}
 \left|V_{kk'}^{(NK)}\right|^2
 \nonumber \\* &\qquad\times
 \delta(E-\varepsilon_k)
 \delta(E+\omega-\varepsilon_{k'})
\end{align}
\end{subequations}
and
\begin{subequations}
\label{Eq13ss}
\begin{align}
 \label{Eq13ssa}
 &\mathbf{\Sigma_P^{<}}(E) = \sum_\alpha\left|V_\alpha^{(P)}\right|^2
 \\* &\quad\times
 \left[\begin{array}{cc}
 (1+N_\alpha)G_{22}^{<}(E+\omega_\alpha) & 0 \\
 0 & N_\alpha G_{11}^{<}(E-\omega_\alpha)
 \end{array}\right] \nonumber \\
 \label{Eq13ssb}
 &\mathbf{\Sigma_P^{>}}(E) = \sum_\alpha\left|V_\alpha^{(P)}\right|^2
 \\* &\quad\times
 \left[\begin{array}{cc}
 N_\alpha G_{22}^{>}(E+\omega_\alpha) & 0 \\
 0 & (1+N_\alpha)G_{11}^{>}(E-\omega_\alpha)
 \end{array}\right]
 \nonumber
\end{align}
\end{subequations}
where $N_\alpha$ is the number of photons in mode $\alpha$. The corresponding 
retarded and advanced SEs are more difficult to calculate from the Langreth 
rules. For simplicity we assume, in the spirit of the wide band approximation, 
that all diagonal components of $\Sigma$ are purely imaginary. Consequently
\begin{equation}
 \label{Eq14ss}
 \mathbf{\Sigma^r}(E) = \left[\mathbf{\Sigma^a}(E)\right]^\dagger =
 \frac{\left[\mathbf{\Sigma^{>}}(E)-\mathbf{\Sigma^{<}}(E)\right]}{2}
 \equiv -\frac{i}{2}\mathbf{\Gamma}
\end{equation}
We use (\ref{Eq14ss}) to get the retarded and advanced GFs according to 
$G^r_{mm'}(E)=\left(G^a_{m'm}(E)\right)^{*}=\left[E-\varepsilon_m-
\Sigma^r_{mm}(E)\right]^{-1}\delta_{mm'}$. $G^{<,>}$ needed in (\ref{Eq6ss})
are then obtained from the Keldysh equation $\mathbf{G^{<,>}}(E)=
\mathbf{G^r}(E)\mathbf{\Sigma^{<,>}}(E)\mathbf{G^a}(E)$. After evaluating 
these matrices we can use Eq.~(\ref{Eq6ss}) with the appropriate self energy 
to calculate the desired flux. We are particularly interested in $I_{em}$
and $I_{sd}$.

With regard to the radiative interactions we consider two models:\\
\textbf{(A)} 
To study the effect of pumping the junction with light we use a model with 
a single photon mode of frequency $\omega_0$ and disregard all other modes, 
as described by Eq.~(\ref{Eq5ss}a). In the semiclassical limit for the 
radiation field and in the RWA we can set $N_0=1$ and take the coupling 
$\hat V_0^{(P)}$ as a product of the local electric field amplitude and the 
molecular transition dipole\cite{9ss}. The charge transfer transition on the
bridge is expressed in this model by taking LUMO bridge level more strongly 
coupled to one lead than to the other. For the HOMO this inequality is assumed
weaker or reversed. The flux of interest is $I_{sd}$ under illumination.\\
\textbf{(B)}
To describe current driven spontaneous emission we use the interaction 
(\ref{Eq5ss}b) with all radiation field modes taken in their vacuum state. 
The frequency resolved emission, $I_{em}'(\omega)=dI_{em}(\omega)/d\omega$ 
is obtained from (\ref{Eq6ss}) using the frequency resolved self energies
obtained from (\ref{Eq13ss})
\begin{subequations}
\label{Eq15ss}
\begin{align}
 \label{Eq15ssa}
 &\mathbf{\Sigma_P^{<}}(E,\omega)=\frac{\gamma_P(\omega)}{2\pi\rho_P(\omega)}
 \left[\begin{array}{cc}G_{22}^{<}(E+\omega)&0\\0&0\end{array}\right]
 \\
 \label{Eq15ssb}
 &\mathbf{\Sigma_P^{>}}(E,\omega)=\frac{\gamma_P(\omega)}{2\pi\rho_P(\omega)}
 \left[\begin{array}{cc}0&0\\0&G_{11}^{>}(E-\omega)\end{array}\right]
\end{align}
\end{subequations}
with the radiative width
\begin{equation}
 \label{Eq16ss}
 \gamma_P(\omega) = 2\pi\sum_\alpha\left|V_\alpha^{(P)}\right|^2
 \delta(\omega-\omega_\alpha)
 = 2\pi\left|V_\alpha^{(P)}\right|^2_\omega \rho_P(\omega)
\end{equation}
and the density of photon modes, $\rho_P(\omega)=\omega^2/(\pi^2c^3)$ 
where $c$ is the speed of light. The total emission flux is found from 
$I_{em}^{tot}=\int_0^\infty d\omega\,I_{em}'(\omega)$.	

\begin{widetext}
In general, the coupled equations for the self energies and the Green functions
have to be solved self consistently. The numerical results displayed below 
were obtained from such an iterative solution. More transparent closed forms 
may be written under the additional simplifying assumptions:
(a) $\varepsilon_{21}$ is large relative to the total widths of levels 1 and 2; 
(b) $C_{NK}(E,\omega)$; $K=L,R$, Eq.~(\ref{Eq12ssd}), is taken constant 
independent of $E$ and $\omega$; (c) $\gamma_P(\omega)$, Eq.~(\ref{Eq16ss}), 
is also assumed constant and (d) keep only teh lowest (second) order in the
radiative coupling $\hat V_P$. It is then possible to express the relevant 
self energies and the resulting currents in terms of the electronic 
populations $n_1$ and $n_2$ of levels 1 and 2 using 
$n_m=(2\pi i)^{-1}\int_{-\infty}^{+\infty}dE G_{mm}^{<}(E)$, $m=1,2$. 
In addition, for case \textbf{(A)} we are interested in the small or no 
voltage regime, whereupon to a good approximation $n_1=1$ and $n_2=0$. 
This leads, for model \textbf{(A)}, to
\begin{align}
 \label{Eq17ss}
 I_{sd} &= \int\frac{dE}{2\pi\hbar}\sum_{m=1,2}\Gamma_{ML,m}G^r_{mm}(E)
 \Gamma_{MR,m}G^a_{mm}(E)
 \left[f_L(E)-f_R(E)\right]
 \\
 &+ \left|V_0^{(P)}\right|^2\int\frac{dE}{2\pi\hbar}
 \frac{\Gamma_{ML,1}\Gamma_{MR,2}f_L(E-\omega_0)[1-f_R(E)]
      -\Gamma_{ML,2}\Gamma_{MR,1}f_R(E-\omega_0)[1-f_L(E)]}
  {\left[(E-\varepsilon_2)^2+(\Gamma_2/2)^2\right]
   \times\left[(E-\omega_0-\varepsilon_1)^2+(\Gamma_1/2)^2\right]}
 \nonumber
\end{align}
The first term on the right is the usual Landauer term that vanishes at zero
potential bias, $f_L=f_R$. The second term shows explicitly the effect of 
pumping. In the absence of bias we get
\begin{equation}
 \label{Eq18ss}
 I_{sd} = \left|V_0^{(P)}\right|^2\int\frac{dE}{2\pi\hbar}
 \frac{f(E-\omega_0)[1-f(E)]
  \left\{\Gamma_{ML,1}\Gamma_{MR,2}-\Gamma_{ML,2}\Gamma_{MR,1}\right\}}
 {\left[(E-\varepsilon_2)^2+(\Gamma_2/2)^2\right]\times
  \left[(E-\omega_0-\varepsilon_1)^2+(\Gamma_1/2)^2\right]}
\end{equation}
where $f=f_L=f_R$. For our choice of parameters below the analytical results
(\ref{Eq17ss}) and (\ref{Eq18ss}) provide excellent approximations to the 
full self-consistent calculation. Furthermore, when $\omega$ is not too far 
from its resonance value $\varepsilon_{21}$ we can approximate
$f(E-\omega_0)[1-f(E)]$ by $1$ to get
\begin{equation}
 \label{Eq19ss}
 I_{sd} = \frac{\left|V_0^{(P)}\right|^2}{\hbar}
 \frac{\Gamma}{(\varepsilon_2-\omega_0-\varepsilon_1)^2+(\Gamma/2)^2}
 \times
 \frac{\Gamma_{ML,1}\Gamma_{MR,2}-\Gamma_{ML,2}\Gamma_{MR,1}}{\Gamma_1\Gamma_2}
\end{equation}
Eqs.~(\ref{Eq18ss}) and (\ref{Eq19ss}) show explicitly how asymmetry in the 
HOMO and LUMO couplings to the metal electrodes leads to photocurrent in the 
present model.

For model~\textbf{(B)} the frequency resolved emission is obtained from
(\ref{Eq6ss}) and (\ref{Eq15ss}) in the form
\begin{equation}
 \label{Eq20ss}
 I_{em}'(\omega) = \frac{\gamma_P(\omega)}{\hbar}\int_{-\infty}^{+\infty}
 \frac{dE}{2\pi}
 \left\{
 \frac{f_L(E+\omega)\Gamma_{ML,2}+f_R(E+\omega)\Gamma_{MR,2}}
 {(E+\omega-\varepsilon_2)^2+(\Gamma_2/2)^2}
 \times
 \frac{[1-f_L(E)]\Gamma_{ML,1}+[1-f_R(E)]\Gamma_{MR,1}}
 {(E-\varepsilon_1)^2+(\Gamma_1/2)^2}\right\}
\end{equation}
\end{widetext}
and the integrated emission is obtained, using the same approximations as 
before, in the anticipated form
\begin{equation}
 \label{Eq21ss}
 I_{em}^{tot} = \frac{\gamma_P(\varepsilon_{21})}{\hbar}n_2(1-n_1) 
\end{equation}
The level populations $n_1$ and $n_2$ should be obtained from the full self 
consistent calculation. For $\Phi>\varepsilon_{21}$ both LUMO and HOMO bridge 
levels are well inside the energy window between the leads' chemical 
potentials. A good approximation for these populations is (written for the case of negatively biased left electrode) $n_2=\Gamma_{ML,2}/\Gamma_2$; 
$n_1=\Gamma_{ML,1}/\Gamma_1+n_2(B_N+\gamma_P)/\Gamma_1$. (\ref{Eq21ss})
then leads to
\begin{equation}
 \label{Eq22ss}
 I_{em}^{tot} = \frac{\gamma_P}{\hbar}
 \frac{\Gamma_{ML,2}\Gamma_{MR,1}}{\Gamma_1\Gamma_2}
\end{equation}
Comparing to (\ref{Eq18ss}) we see that tailoring the molecule-leads coupling
asymmetry such that resonance radiation will induce electron flow in 
a given direction enhances light emission under such bias that leads to 
electron current in the opposite direction.

\begin{figure}[htbp]
\includegraphics[width=\columnwidth]{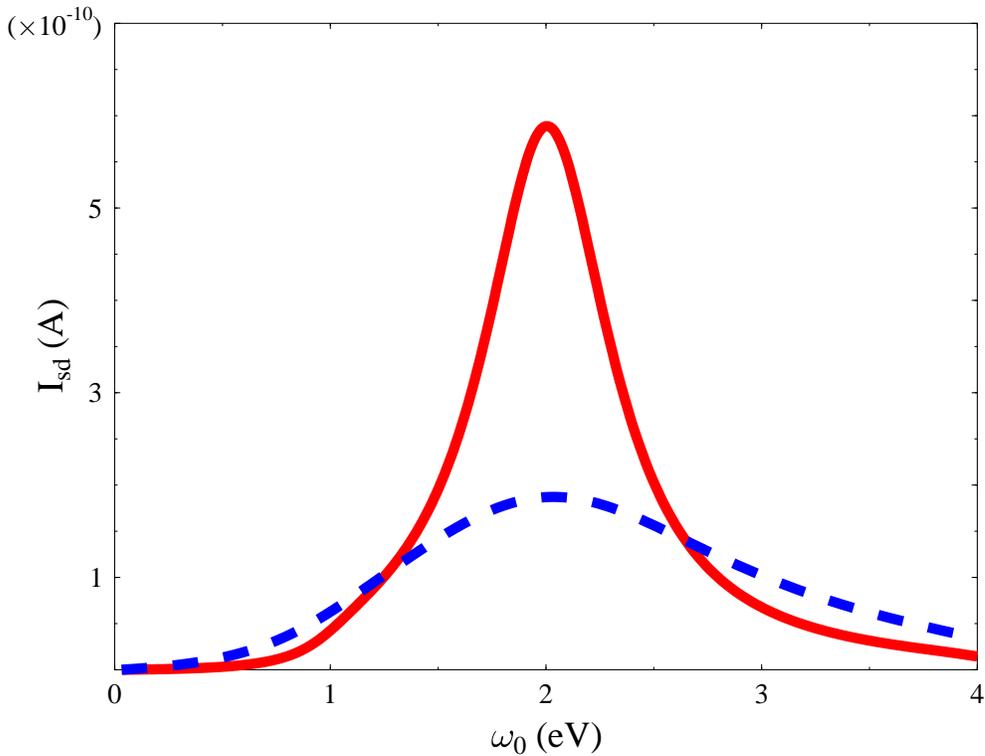}
\caption{\label{Fig2}The light induced current in the model of Fig.~\ref{Fig1}
in the absence of external potential bias. Full (red) line:
$T=300$~K, $\varepsilon_{21}=2$~eV, $\Gamma_{ML,1}=\Gamma_{MR,1}=0.2$~eV, 
$\Gamma_{ML,2}=0.02$~eV, $\Gamma_{MR,2}=0.3$~eV, $\gamma_P=10^{-6}$~eV, 
$B_{NL}=B_{NR}=0.1$~eV and $V_0^{(P)}=10^{-3}$~eV. Dashed (blue) line ---
same parameters except $\Gamma_{MK,m}\times 3$ ($m=1,2$) is used.}
\end{figure}

\begin{figure}[htbp]
\includegraphics[width=\columnwidth]{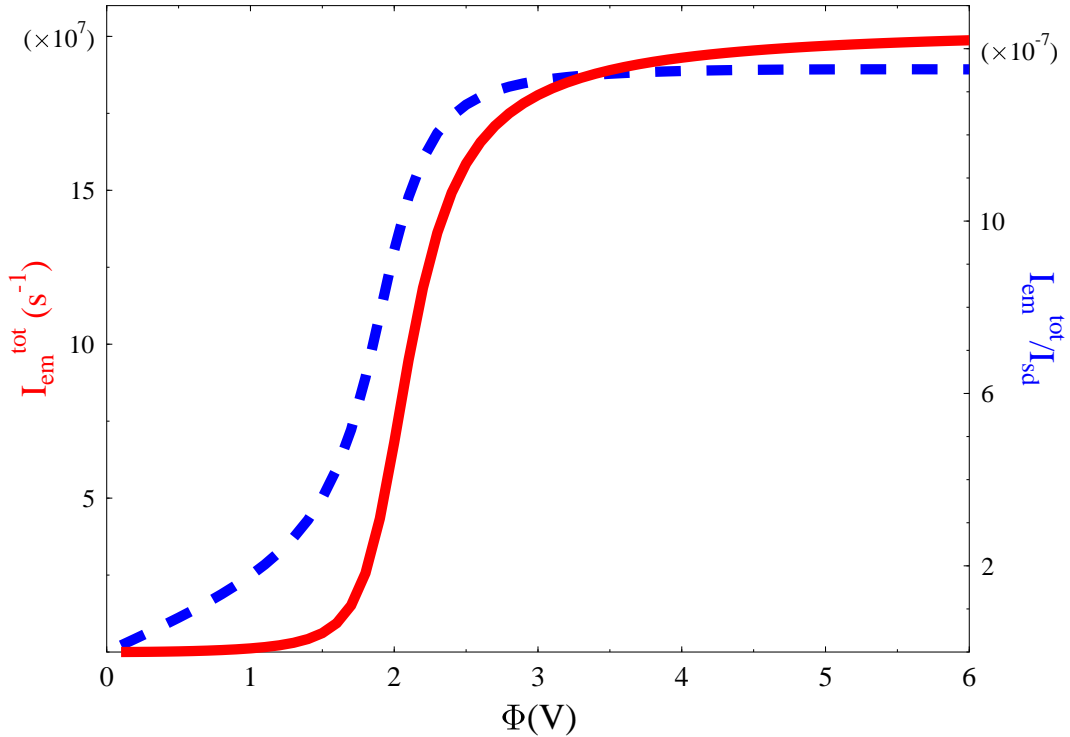}
\caption{\label{Fig3}The integrated photon emission rate (full line, red)
and yield (dashed line, blue) plotted against the bias voltage. Parameters
are as in Fig.~\ref{Fig2} except $\Gamma_{MK,m}=0.1$~eV, $K=L,R$, $m=1,2$.}
\end{figure}

Numerical results obtained from the self consistent calculation are shown in 
Figs.~\ref{Fig2} and \ref{Fig3}. Figure~\ref{Fig2} shows the light induced
current obtained using a set of reasonable junction parameters. In particular,
for a molecule with transition dipole moment $\sim 1$D the choice 
$V_0^{(P)}=10^{-3}$~eV corresponds to an incident field intensity 
$\sim 10^8$~watt/cm${}^2$.

Fig.~\ref{Fig3} shows the integrated emission as well as the predicted photon
emission yield. The later is defined as the ratio between the emitted photon 
current and the electronic current passed by the junction --- both displayed 
as functions of the bias voltage.

These results, obtained under rather conservative choices of damping 
parameters, indicate that both light induced current and the 
molecular mechanism (as opposed to the surface plasmon mechanism) for current 
induced emission in tunnel junctions can lead to measurable signals. 
This molecular mechanism can in principle be distinguished from the plasmon 
mechanism for light emission by the frequency dependence of the resolved 
emission, as implied by Eq.~(\ref{Eq20ss}). 

In conclusion, we have described a model that accounts for observed current 
induced light emission from molecular tunnel junctions and provides the tools 
for determining the intensity and yield of such emission as functions of key 
junction parameters. The same model predict that resonant light induced current
can occur in junctions employing molecular bridges with strong charge-transfer 
transition to their first excited state. We have concluded, using reasonable 
parameter, that light driven electronic currents and current driven light 
emission are realistic possibilities. The linewidths and lineshapes associated 
with these signal can also be analyzed by the same theoretical framework. 
Correlating observations with predictions made in this paper should help 
the interpretation of future experiments in this direction.

\begin{acknowledgments}
We acknowledge support from the Israel Science Foundation the US-Israel 
BSF and by the NSF/NCN project through Northwestern University. We thank 
Profs. T.~Marks and M.~Ratner for helpful discussions and support.
\end{acknowledgments}


\begin{thebibliography}{99}
\bibitem{9ss}In disregarding all $\alpha \neq 0$ modes we neglect radiative
             damping relative to the other relaxation channels present.
\end{thebibliography}
\end{document}